# Scaling Properties of Critical Phase Change Conditions in $Ge_2Sb_2Te_5$ Nanopillars

O. Ozatay, B. Stipe, J. A. Katine, and B. D. Terris
*Hitachi Global Storage Technologies-San Jose Research Center*
*San Jose CA 95135*

*ABSTRACT*

We have measured the critical phase change conditions induced by electrical pulses in $Ge_2Sb_2Te_5$ nanopillar phase change memory devices by constructing a comprehensive resistance map as a function of pulse parameters (width, amplitude and trailing edge). Our measurements reveal that the cooling scheme and the details of the contact geometry play the dominant role in determining the final phase composition of the device. A three-dimensional finite element model of the electro-thermal physics not only provides good insights into the underlying physical mechanisms of the switching dynamics but also quantitatively accounts for the observed scaling behaviour.

Following the original work of Ovshinsky[1], the reversible switching phenomena observed in disordered semiconductors has attracted substantial theoretical[2,3,4,5,6,7] and experimental[8,9,10,11,12,13] efforts targeted at gaining a deep understanding of the physical principles that govern the switching dynamics, and also learning how to control and optimize the power requirements coupled with characteristic switching timescales. Dramatic and ultra-fast (nanosecond scale) changes in the physical properties such as electrical resistivity and optical reflectivity upon amorphization or crystallization makes chalcogenides ideal potential candidates for a universal non-volatile memory device, especially when the bulk thin film switching characteristics are highly scalable to nanometer dimensions. The principle of phase-dependent, large changes in the optical reflectivity induced by local heating through laser pulses has already been exploited in commercial optical storage media such as DVDs. On the other hand, the practical utilization of three to four orders of magnitude changes in the electrical resistivity induced by local joule heating through electrical pulses necessitates a detailed experimental characterization and theoretical modeling of the critical phase change conditions and their scaling properties.

In this work we use a resistance mapping technique whereby we apply electrical pulses of various amplitudes, widths and trailing edges and deduce the resulting phase composition in nanoscale $Ge_2Sb_2Te_5$ nanopillar devices. Such resistance map measurements, when repeated for devices with different contact sizes and geometries, provide a detailed picture of the scaling behaviour of the critical phase change conditions and the associated switching dynamics. A comparison of the observed characteristics with a three-dimensional finite element model (3D-FEM) of the electro-thermal physics enables a good assessment of the resulting device performance while providing physical insights into the nature of resistance switching behaviour.

The device structure under study consists of a 50nm $Ge_2Sb_2Te_5$ (GST) chalcogenide phase change material sandwiched in between two 50 nm $W_{0.8}Ti_{0.2}$ (WTi) heating element layers. A schematic drawing is shown in Fig.1 (a). The top WTi electrode is patterned into a nanopillar by using advanced e-beam lithography and ion-mill etching techniques that stop within a nm of the top GST-WTi interface utilizing a secondary-ion-mass-spectroscopy (SIMS) signal to avoid ion-mill damage on the phase change material in an attempt to preserve bulk film properties. The electrical contacts to the device are made through Ta (5nm)/Au (~175nm) bilayers. All the devices are isolated by an alumina field-oxide layer. The active phase transition region is defined by the top GST-WTi interface where the local concentration of the applied current enables effective heating above the melting point whereas both the top and bottom metallic electrodes act as efficient heat sinks to quickly absorb the heat during cooling. Fig. 1 (b) shows a cross-sectional scanning electron microscope (SEM) image of a 75 nm diameter nanopillar. It is clearly observable that the actual contact area at the top GST-WTi interface can be 20-30% larger than the pillar diameter due to tapering as a result of ion-mill re-deposition effects.

We have fabricated devices with linear dimensions in the range of 75 nm to 1 μm and with square and circular pillar contact geometries. Fig. 2 (a) shows the current-voltage (IV) characteristics of both amorphous (in blue) and crystalline (in red) states of a 600 nm square device. The crystalline (SET) state exhibits a low resistance (~90Ω) accompanied by ohmic IV characteristics whereas the amorphous (RESET) state resistance at low bias (<~0.6V) is of the order of 10s of kΩs. Increasing the voltage bias above a threshold value (~0.65V in this case) a snap-back occurs as a result of formation of highly conductive paths in the amorphous material (also known as Ovonic threshold switching (OTS)). Since, after the snap-back, the material is still in the amorphous state this effect is believed to be of electronic origin (not the result of a structural change). While still not very well understood, it has been attributed to be due to a competition between impact ionization and recombination via valence alternation pairs[6], thermal instabilities[14] or Poole-Frenkel trap-limited conduction[15]. After the snap-back, which allows relatively high currents to flow, substantial joule heating causes an accelerated transformation to the thermodynamically favorable crystalline (SET) state via Ovonic memory switching (OMS).

The reset operation is performed by applying pulse amplitudes large enough for the GST to be locally heated above its melting point (~620 °C) and then super-cooling (melt-quenching) it below its glass transition temperature (~300 °C) into the amorphous phase. The switching back to the crystalline state can be achieved by sweeping the dc bias beyond the snap-back voltage as described above or alternatively by applying pulses of amplitude insufficient to melt but above the threshold for triggering crystal nucleation and growth. The former method requires the use of a set current compliance to protect the smallest devices, since excessive heating after OTS can result in device failure. Five such consecutive switching events following a reset pulse are shown in Figs. 2(b) and (d) corresponding to a 90nm square and a 75 nm diameter circular contact respectively. The presence of parabolic (non-ohmic) IV characteristics in the set state (red curves) implies incomplete crystallization (i.e. residual amorphous content). Inhomogeneities in the phase composition can be visualized as an amorphous matrix in a crystalline background, which during the reset pulse, will only be influenced by heat diffusion from the surroundings, potentially resulting in temperatures not exceeding the melting point. The resulting fluctuations in the initial phase composition and resistance lead to variations in the snap-back voltage from one IV sweep to another. Figures 2 (b) and (d) suggest that this effect is more dominant for a square contact than for a circular contact, since the sharp corners in that case lead to current-crowding effects[16] and locally inhomogeneous distributions of temperature gradients (see the discussion below on finite-element calculations). Figure 2 (c) shows the corresponding pulsed reset operations (resistance versus pulse amplitude) of the 75 nm diameter circular device with 50 ns wide/ 2ns trailing edge pulses. The observation of a reset voltage almost half the snap-back voltage well-justifies the choice of using a current compliance in the IV sweeps since in this case clearly the snap-back process immediately leads to large enough currents to cause local melting of the GST if there is no current limiting mechanism in the measurement setup.

Figure 3 displays the results of a resistance map characterization of the pulsed switching (dc resistance as a function of applied pulse width and amplitude at a given pulse trailing edge) of a 75 nm circular device. Prior to each pulse application, the device is prepared in the predominantly crystalline ($C_A$), amorphous ($A_C$) states for the reset (Fig.3 (a),(b) and (c)) and set (Fig.3 (d), (e) and (f)) operations respectively. The standard crystallization pulse used was 2V in amplitude, 1 µs wide with a 150 ns trailing edge whereas the standard amorphization pulse was 0.8V in amplitude, 50 ns wide with a 2ns trailing edge. As can be observed in the resistance maps in Fig. 3, the snap-back voltage (~0.85V) appears to be higher than the threshold voltage for the reset operation (~0.6V). Therefore, analogous to the current compliance used in crystallization by a dc sweep, in this case we used a 1 kΩ series-load resistor with the GST device to avoid damage due to excess currents. The lack of impedance matching between the load resistor and the output of the pulser (50Ω) leads to a 1/10 attenuation in the transmitted pulse amplitude with no detectable pulse distortion. Further multiple pulse reflections/distortions between the GST device and the 1kΩ resistor were avoided by placing them in close proximity (within 3 cm). All of the reported pulse parameters correspond to the actual values that appear on the GST device.

The reset operation resistance map in Fig. 3(a), (b) and (c) suggest that the pulse trailing edge (i.e. the cooling rate) plays a key role in determining the critical pulse widths needed to achieve amorphization[13]. Increasing the trailing edge from 2 ns to 42 ns (decreasing the cooling rate from ~3 * $10^{11}$ K/s to ~1.5 * $10^{10}$ K/s) results in an increase in the critical pulse width from 10 ns to 60 ns, above which the reset voltage (~0.6V) becomes virtually independent of the applied pulse width. However there exists a narrow range (about 5 ns) below the cited thresholds where the reset operation appears to be possible at the expense of a ~10% increase in the necessary reset pulse amplitude. The set resistance (in the $C_A$ state) was ~10kΩ, suggesting a threshold reset current of ~60 μA.

Figures 3 (d), (e) and (f) show the corresponding resistance maps for the set operation. The set resistance obtained in pulsed switching operation was found to be larger than the resistance of a fresh sample (~1kΩ), which indicated that the pulsed set operation always resulted in a predominantly crystalline state with a residual amorphous content ($C_A$ state). After pulsed switching, a full recovery of the initial resistance (fully crystallized state) was possible in devices with linear dimensions above 100nm by doing a set operation with dc sweeps (sweeping speed ~0.1V/s and without current compliance) suggesting that the pulse induced changes in the set resistance are reversible. This implies that although from Fig. 3 (d) the characteristic crystal nucleation and growth time appears to be of the order of 50 ns, such a time scale is not enough to achieve full crystallization of the GST layer. Increasing the trailing edge of the set pulse from 2 ns to 252 ns results in an increase in the critical pulse widths from ~50 ns to ~250 ns. The critical pulse width in the former measurement is determined by the characteristic crystallization time, whereas in the latter case it is determined by the transition from a triangular pulse (when the pulse width is less than or comparable to the trailing edge) to a rectangular pulse. The pulse amplitude in a triangular pulse is substantially attenuated by a factor equal to the pulse width divided by (2 * trailing edge). Therefore the unsuccessful set operations in those regions imply that the pulse amplitude on the GST actually never reaches the snap-back voltage level.

The results of a 3-D FEM model of the local heating induced by electrical pulses are shown in Fig. 4. This model takes into account the temperature dependence of the thermal conductivities[17,18] and the electrical resistivities[4,18] of both the GST and the WTi layer. The electrical model involves solving the Laplace equation assuming that the electrical conductivity of the GST has no field dependence[4]. The calculated power dissipation acts as the heat source in the thermal model. The calculated temperature of each cell, as determined by the solutions of the heat diffusion equation modifies the electrical resistivity and the thermal conductivity accordingly for the next iteration. The top and bottom electrode surfaces have a fixed thermal boundary condition of T=300K since they act as heat sinks in the cooling process whereas the electrical problem assumes the bottom surface to be grounded with the top electrode surface acting as the electrical pulse source. We note that this simple model does not take into account phase change kinetics (homogeneous-heterogeneous nucleation theories and electrical percolation effects) which are accounted for in a more sophisticated model[4]. Nevertheless the calculated threshold current level scaling for the reset operation shows very good agreement with experimentally determined values as shown in Fig. 4 (c).

Figure 4 (a) shows the temporal evolution of the normalized thermal gradient at the GST-WTi interface for a circular and square contact with 100 nm linear dimensions. During the application of a reset pulse, the local heating pattern clearly shows an inhomogeneous heat distribution in a square contact compared to a circular contact, with preferential local heating at the sharp corners[16]. The inhomogeneous heating scheme in the square contact facilitates the formation of a residual amorphous matrix in the set state leading to more pronounced variations in the snap-back voltage as noted in Fig. 2 (b). The resulting temperature distribution for a 100 nm circular contact at the end of a 50 ns reset pulse is shown in Fig. 4 (b). Since the GST-WTi interface is close to a metal heat sink, the actual hot spot occurs almost half way into the GST layer. If the device is quenched to below its glass transition temperature starting from this thermal profile, it results in an amorphous hemisphere surrounded by crystalline GST and, therefore, a slight increase in the device resistance. If the reset voltage level is below the snap-back threshold as is the case for a 75 nm circular device, the highly insulating-amorphous hemisphere allows more concentrated current flow at its surroundings and therefore an accelerated reset switching (amorphization of the GST-WTi interface).

A further experimental proof for the mixed phase composition comes from the surprising trailing edge dependence of the set resistance as shown in the inset of Fig. 4 (c). As long as the pulse amplitude and pulse width are large enough to trigger crystallization, one would expect that the trailing edge (provided it is above the characteristic nucleation time) should not play a role in the set operation. However, applying 1.7 V amplitude, 0.6 µs set pulses with 50 ns to 250 ns trailing edges results in a change in the set resistance from 20 kΩ to 1kΩ for the 75 nm circle. The temperature distribution shown in Fig. 4(b) suggests that as the device is heated, the crystal nucleation is expected to start from the middle hot spot analogous to the reset operation. However, crystallization of the GST-WTi interface leads to re-melting of the middle hot spot (due to the temperature gradient) which will partially quench into the amorphous phase depending on the trailing edge ( analogous to the reset switching behaviour as shown in Figs. 3(a),(b) and (c)).

In summary, we have employed a resistance mapping technique to probe the critical phase change conditions of GST phase change memory devices with various contact sizes. The scaling behaviour of the reset conditions shows very good agreement with a 3-D FEM model of the electro-thermal physics. Both experimental results and the simulations suggest that the pulsed set and reset operations result in a mixed phase (with a predominantly crystalline or amorphous content) where the amorphous/crystalline volume ratio is highly sensitive to the details of contact geometry and can also be controlled by tuning the cooling rate (pulse trailing edge).

## Figure Captions

**Fig 1. (color online)**

**(a)** Schematic drawing of the GST nanopillar device with WTi top and bottom electrodes.

**(b)** Cross-sectional SEM micrograph of a 75nm diameter nanopillar disk.

**Fig 2: (color online)**

**(a)** IV characteristics of a 600 x 600 nm$^2$ device with blue curve corresponding to the amorphous (RESET) state displaying threshold switching (OTS) and memory switching (OMS) to the crystalline state, and the red curve corresponding to the crystalline (SET) state.

**(b),(d)** Five consecutive IV measurements of a 90nm square and 75 nm circular pillar respectively with each measurement starting from the amorphous state (following a pulsed reset operation)

**(c)** 75 nm circular device resistance versus pulse amplitude for 50ns wide voltage pulses with a 2 ns trailing edge displaying the 5 consecutive reset operations before each IV measurement shown in (d).

**Fig. 3: (color online)**

Resistance maps of the reset operation of a 75 nm diameter circular pillar as a function of pulse width and amplitude for 2 ns **(a)**, 22 ns **(b)** and 42 ns **(c)** trailing edges and of the set operation with 2 ns **(d)**, 152 ns **(e)**, and 252 ns **(f)** trailing edges.

**Fig. 4: (color online)**

**(a)** Temporal evolution of the simulated temperature gradient profile of a 100nm diameter circular contact (top) and a square contact with 100 nm on a side (bottom) (at 40 ns and 50 ns during a 50ns reset pulse).

**(b)** Cross-sectional view of the simulated spatial temperature distribution in a 100 nm diameter contact at the end of a 50ns reset pulse.

**(c)** Reset current scaling with contact area red experimental data and blue simulation results with the solid lines being linear fits. (Inset) Set resistance variation with pulse trailing edge for a pulse level of 1.7 V and a pulse width of 0.6 µs for a 75 nm circular contact.

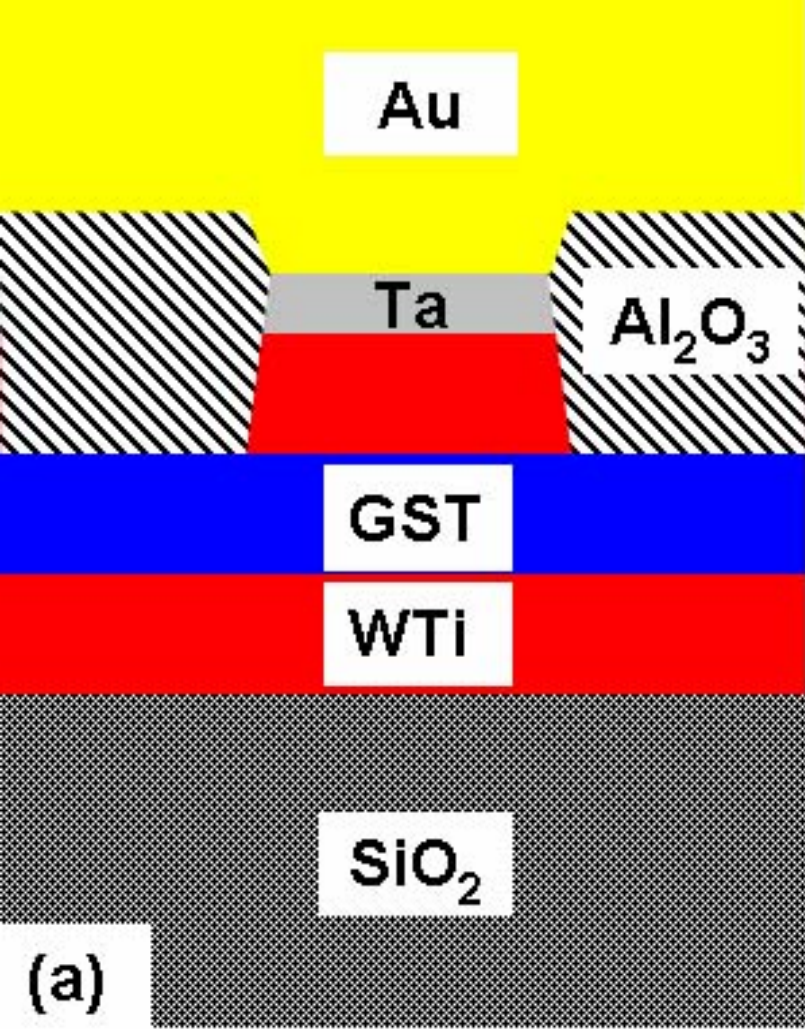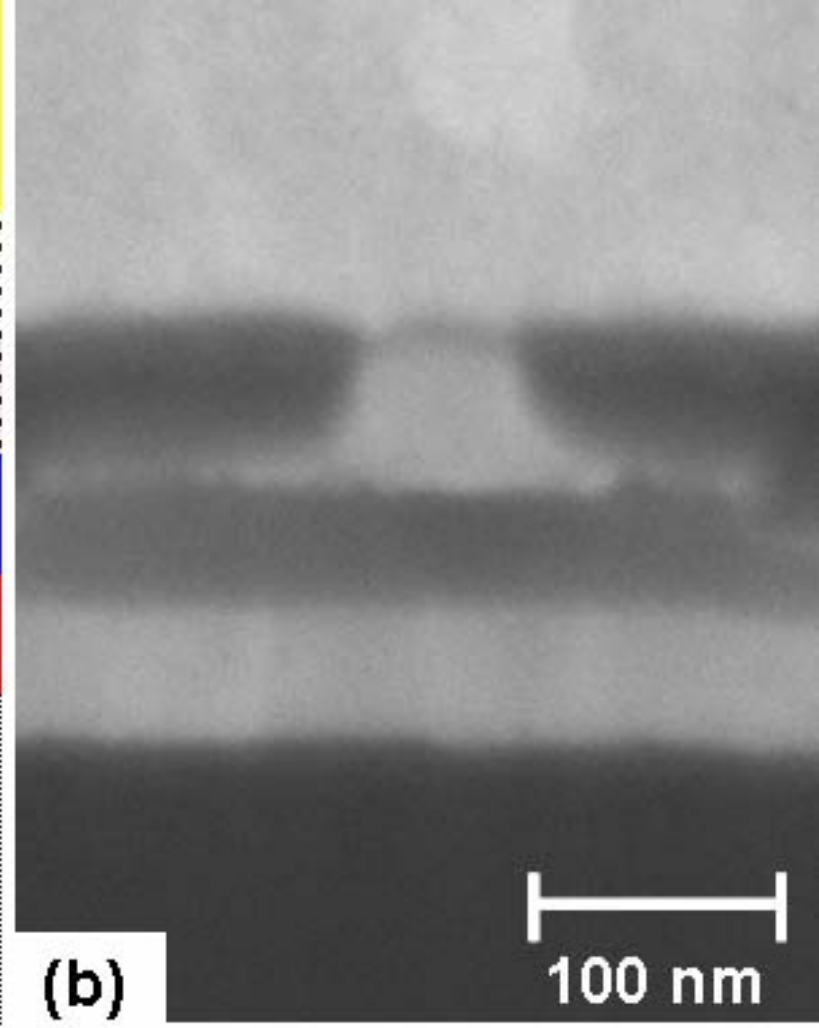

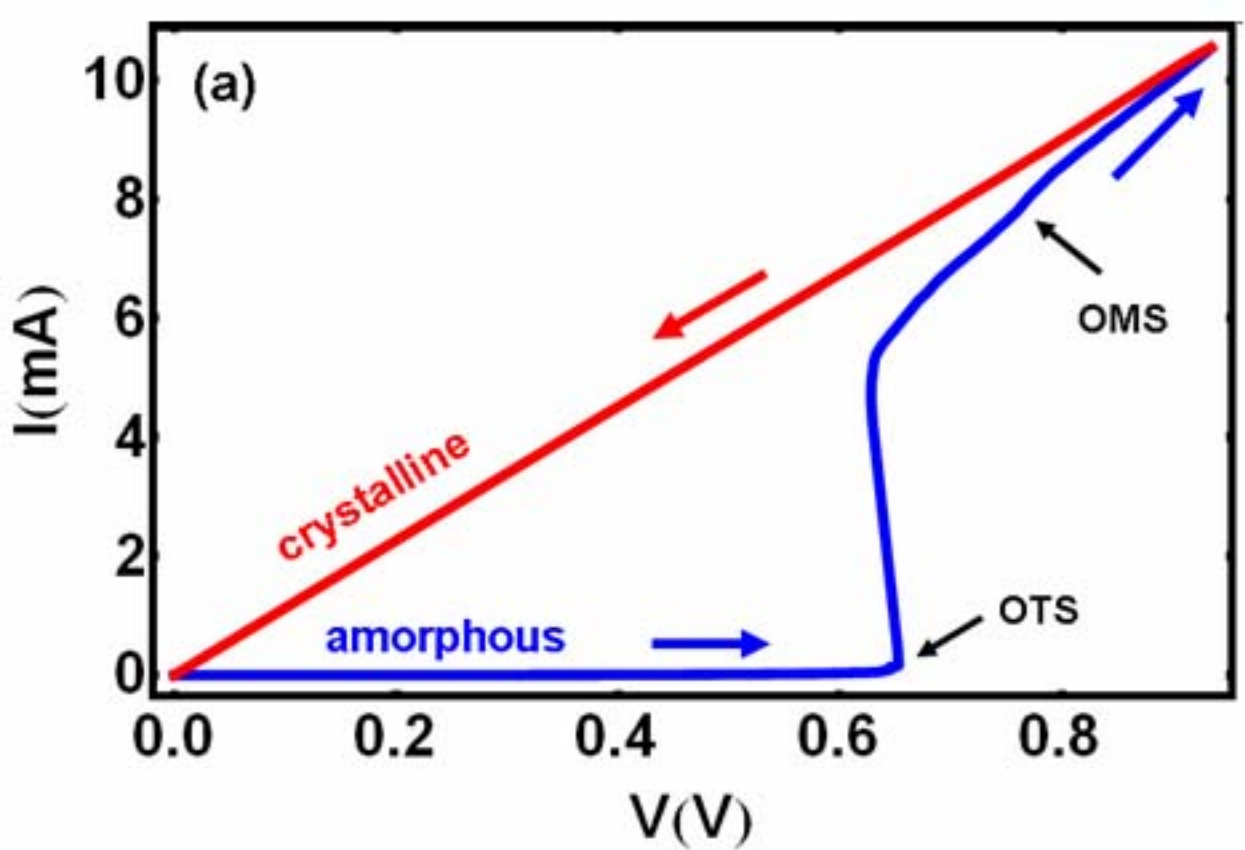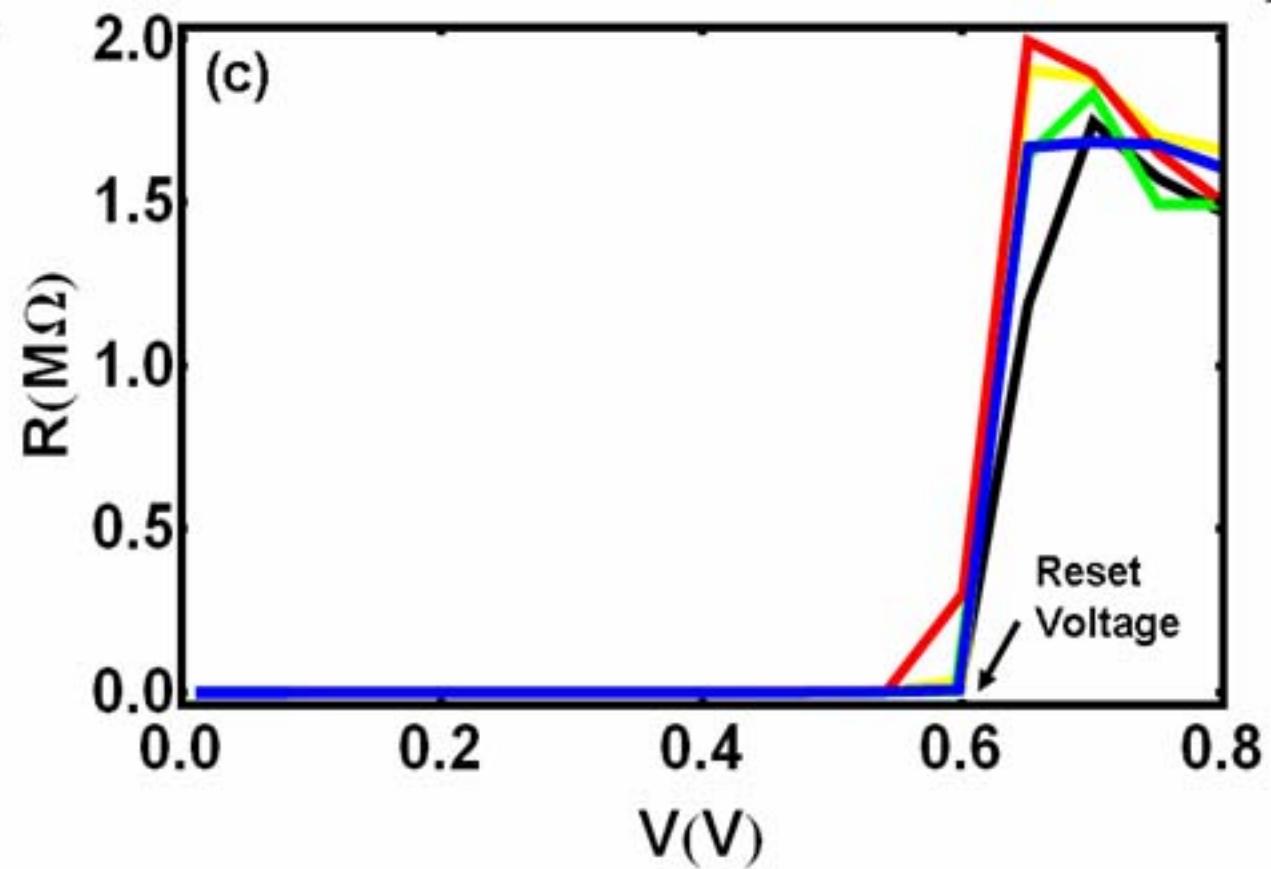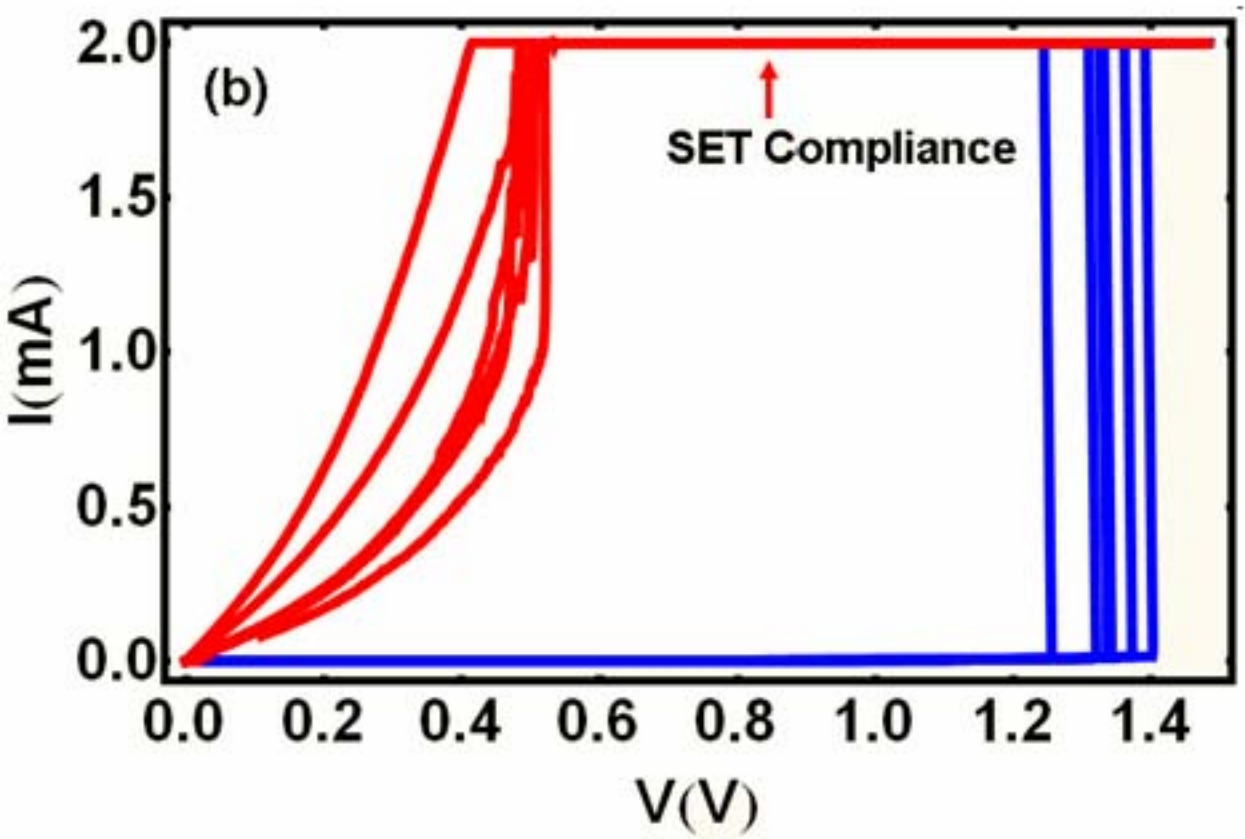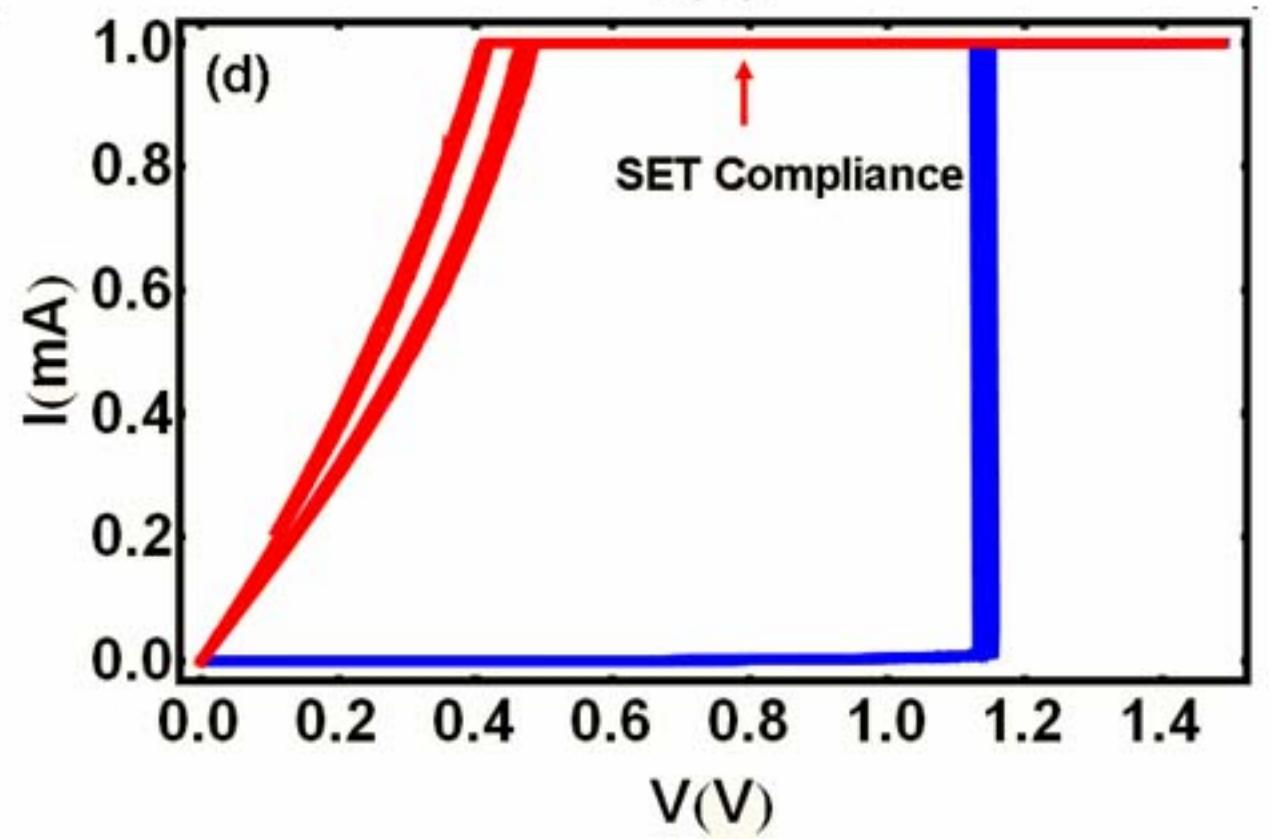

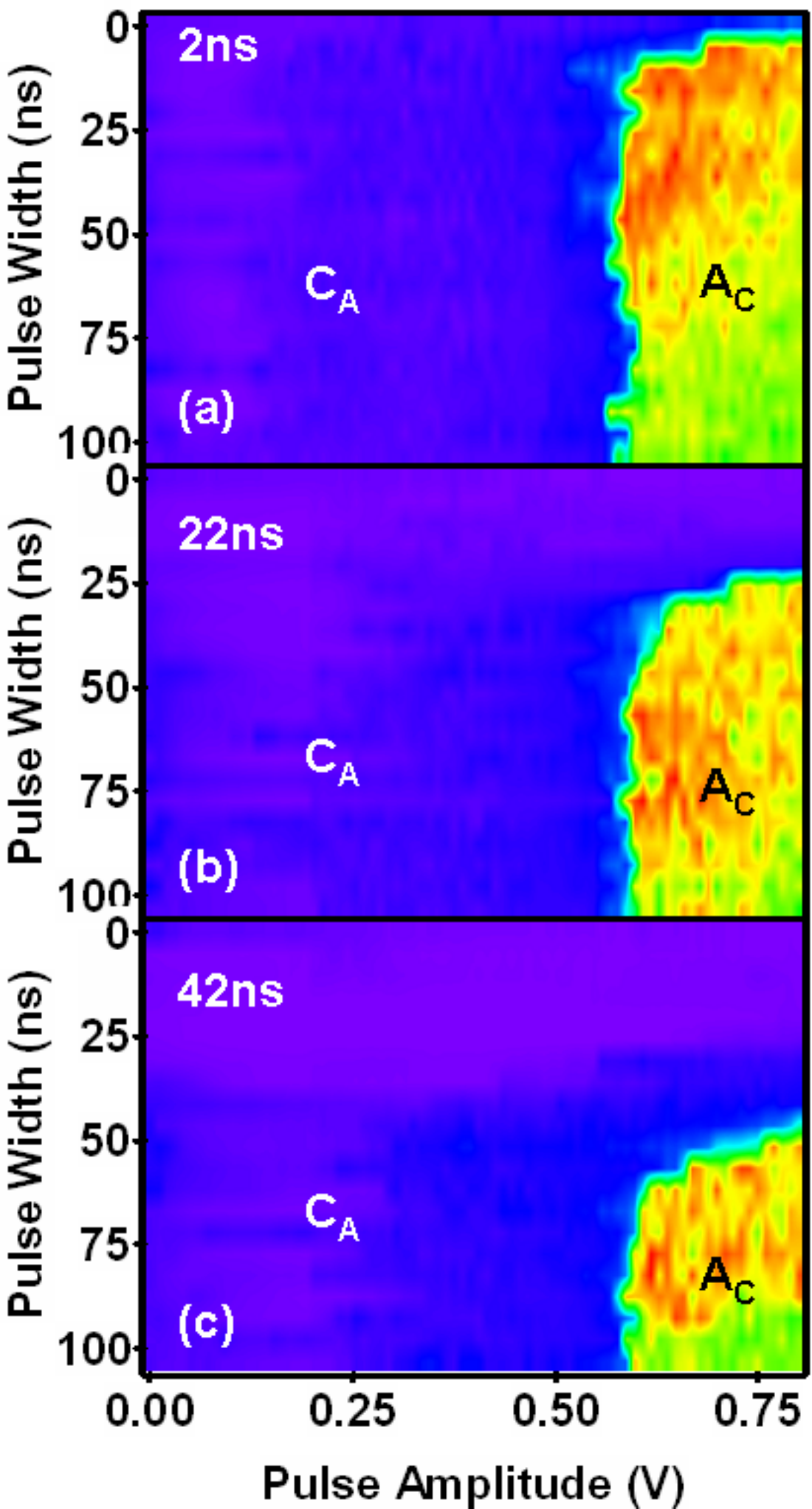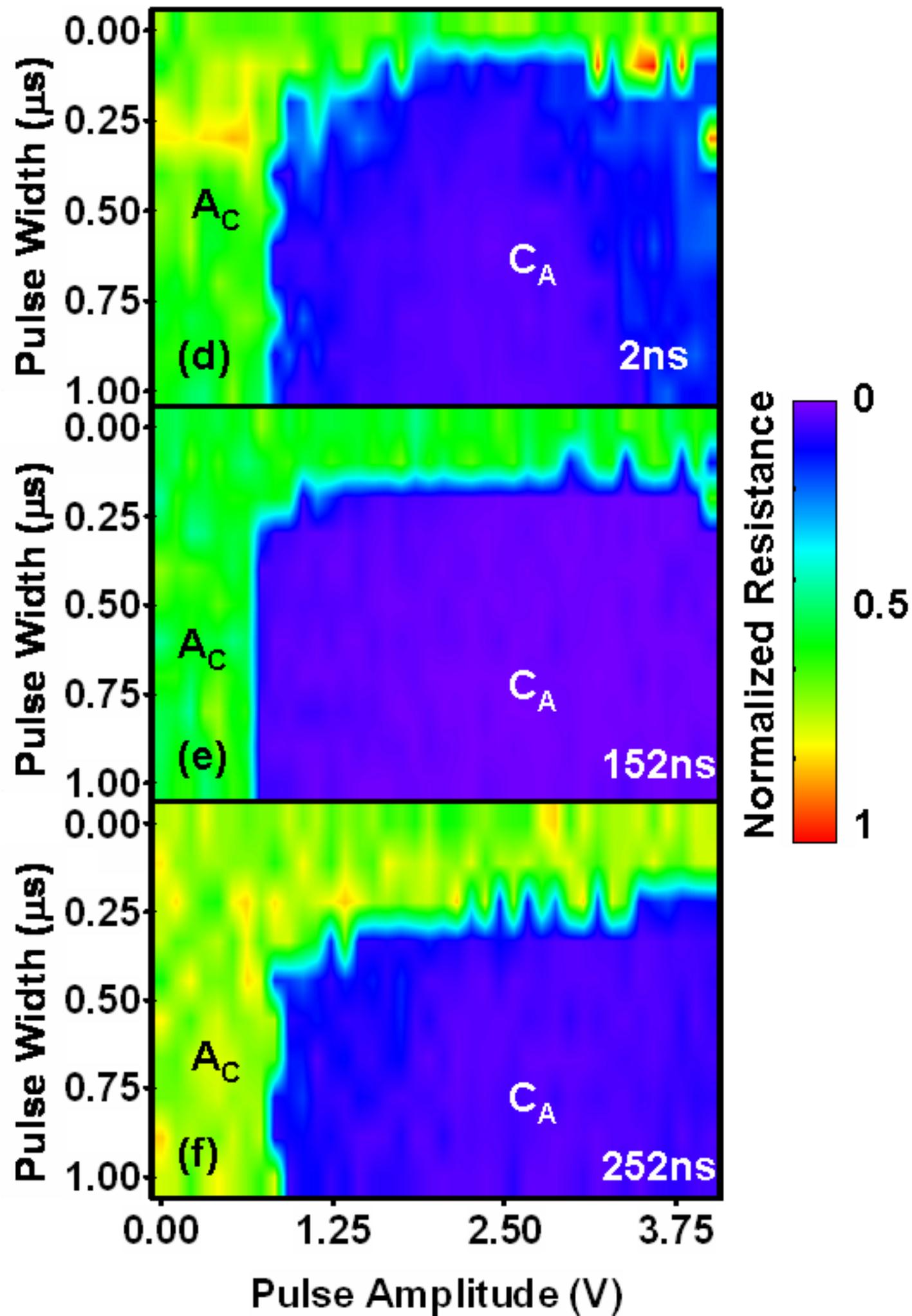

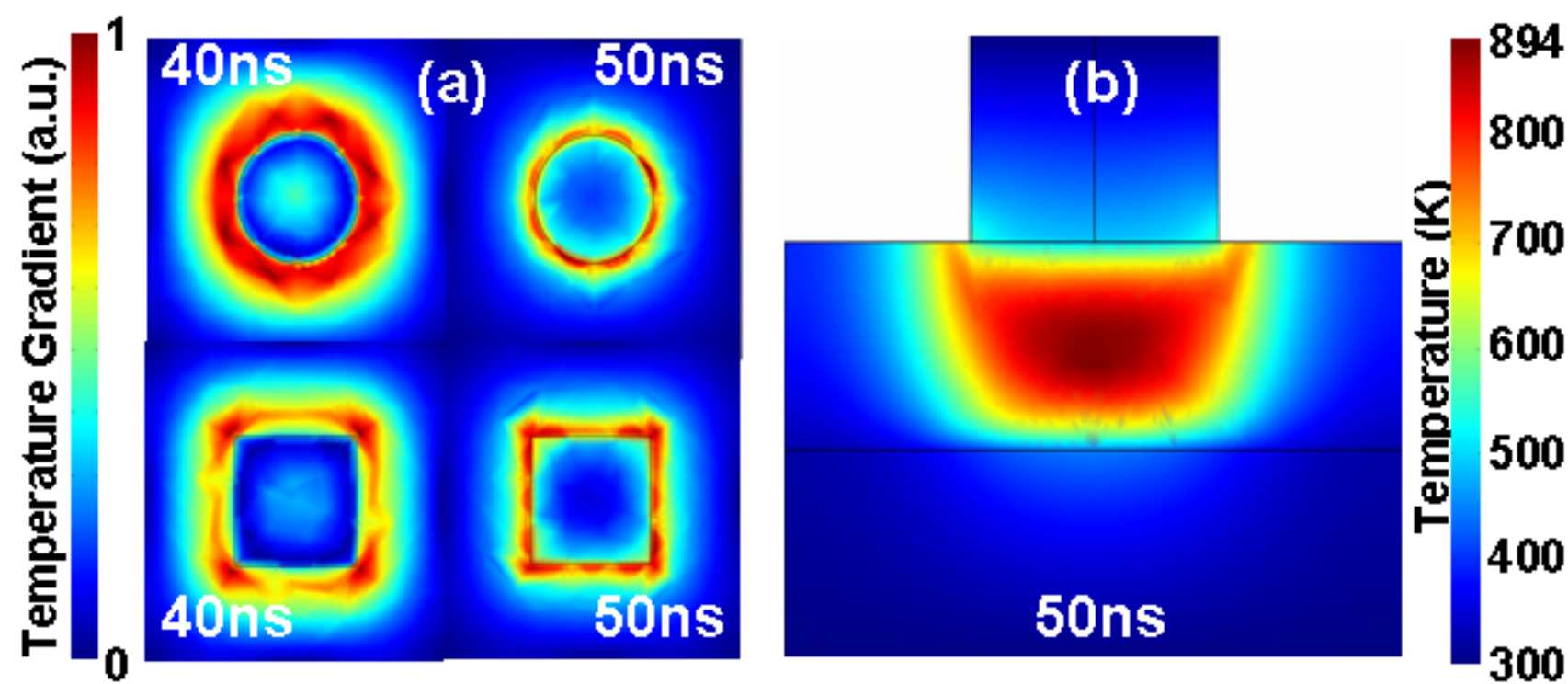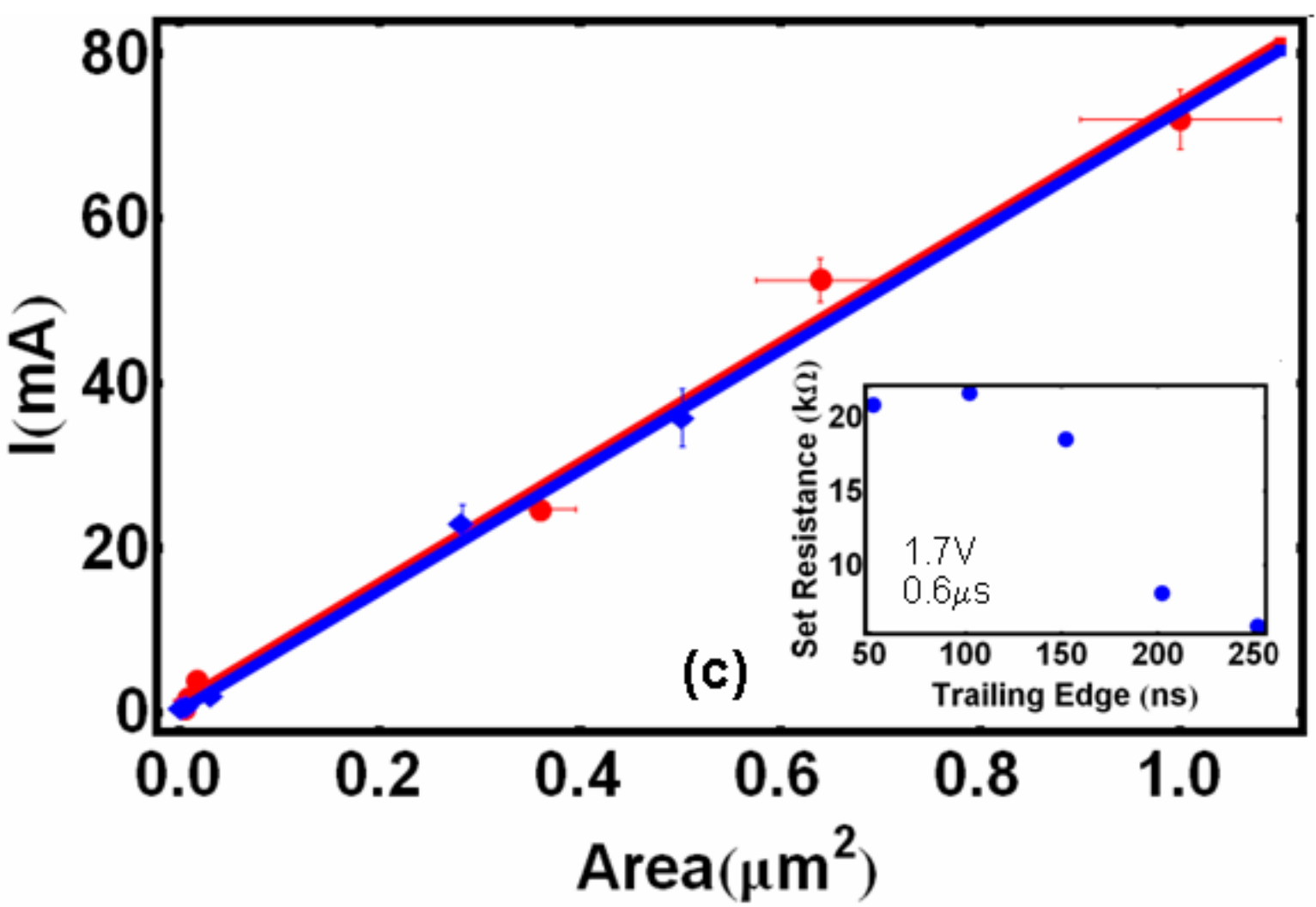